\documentclass[%
 reprint,
 amsmath,amssymb,
 aps,
]{revtex4-1}

\usepackage{graphicx}
\usepackage{dcolumn}
\usepackage{bm}
\usepackage{textcomp}
\usepackage{xcolor}
\usepackage{siunitx}
\usepackage{tikz}
\usepackage{placeins}

\newcommand{\bE}{\mathbf{E}}

\begin{document}

\title{Superconducting Niobium Tip Electron Beam Source}
 
\author{C.W. Johnson$^1$, A.K. Schmid$^1$, M. Mankos$^2$, R. R\"{o}pke$^3$, N. Kerker$^3$, I.S.~Hwang$^5$, E.K. Wong$^1$, D.F. Ogletree$^1$, A.M. Minor$^{1,4}$ and A. Stibor$^{1,3*}$}
\affiliation{\mbox{$^1$Lawrence Berkeley National Lab, Molecular Foundry, Berkeley, USA, *email: astibor@lbl.gov}
\mbox{$^2$Electron Optica Inc., Palo Alto, USA}\\
\mbox{$^3$Institute of Physics and LISA$^+$, University of T\"{u}bingen, T\"{u}bingen, Germany} 
\mbox{$^4$Department of Materials Science and Engineering, University of
California, Berkeley, USA}
\mbox{$^5$Institute of Physics, Academia Sinica, Nankang, Taipei, Taiwan, Republic of China}}

\begin{abstract}
Modern electron microscopy and spectroscopy is a key technology for studying the structure and composition of quantum and biological materials in fundamental and applied sciences. High-resolution spectroscopic techniques and aberration-corrected microscopes are often limited by the relatively large energy distribution of currently available beam sources. This can be improved by a monochromator, with the significant drawback of losing most of the beam current. Here, we study the field emission properties of a monocrystalline niobium tip electron field emitter at 5.2 K, well below the superconducting transition temperature. The emitter fabrication process can generate two tip configurations, with or without a nano-protrusion at the apex, strongly influencing the field-emission energy distribution. The geometry without the nano-protrusion has a high beam current, long-term stability, and an energy width of around \SI{100}{meV}. The beam current can be increased by two orders of magnitude by xenon gas adsorption. We also studied the emitter performance up to \SI{82}{K} and demonstrated the beam's energy width can be below \SI{40}{meV} even at liquid nitrogen cooling temperatures when an apex nano-protrusion is present. Furthermore, the spatial and temporal electron-electron correlations of the field emission are studied at normal and superconducting temperatures and the influence of Nottingham heating is discussed. This new monochromatic source will allow unprecedented accuracy and resolution in electron microscopy, spectroscopy, and high-coherence quantum applications. 

\noindent 
\end{abstract}

\maketitle
\setlength{\parindent}{5mm}

\noindent {\bf \large INTRODUCTION}\\
Electron beam field emitters are foundational in modern electron-optical applications. The performance of electron microscopes \cite{chen2020imaging,ophus2019four}, interferometers \cite{turner_interaction-free_2021,kerker_quantum_2020,rembold_vibrational_2017}, sensors \cite{pooch_compact_2017} and quantum information science applications \cite{ropke_data_2021} rely on intense, stable, coherent and monochromatic beam sources with high brightness. The development of novel beam sources exploiting their nanoscopic quantum environment is opening new areas in microscopy, such as laser-pulsed tip emitters \cite{feist_ultrafast_2017,ehberger_highly_2015} allowing pulse-probe microscopy of dynamic behavior on the nanosecond scale, the single-atom tips \cite{kuo_noble_2006,chang_fully_2009,Schuetz2014} for matter-wave experiments with high coherence or carbon nanotube field emitters \cite{de2002high} with high brightness. Recently, a micro-engineered LaB$_6$ nanowire-based electron source with an on-tip integrated passive collimator achieved atomic resolution in a transmission electron microscope \cite{zhang2021high} and a highly monochromatic cold flat single-crystal Cu(100) surface source based on near-threshold photoemission was reported \cite{karkare_ultracold_2020}. However, commercialized microscope hot Schottky field emitters have typically energy widths of $\Delta E \approx $ \SI{750}{meV} and “cold" (meaning room temperature) field emitters have $\Delta E  \approx $ \SI{300}{meV}. This large energy distribution limits current state-of-the-art analytical methods in electron microscopy, such as high-resolution vibrational spectroscopy \cite{krivanek2014vibrational} or surface-sensitive imaging techniques like low energy electron microscopy (LEEM) \cite{bauer2014basic}. The energy distribution can be reduced down to \SI{9}{\meV} \cite{krivanek2014vibrational} by the application of monochromators \cite{mook2000construction}, but this removes most of the beam current and results in significantly extended measurement times, potentially causing problems with sample stability. There is a need for stable, bright, and coherent electron emitters with intrinsically narrow energy distributions.  

\begin{figure*}[t]
\centering
\includegraphics[width=1.0\textwidth]{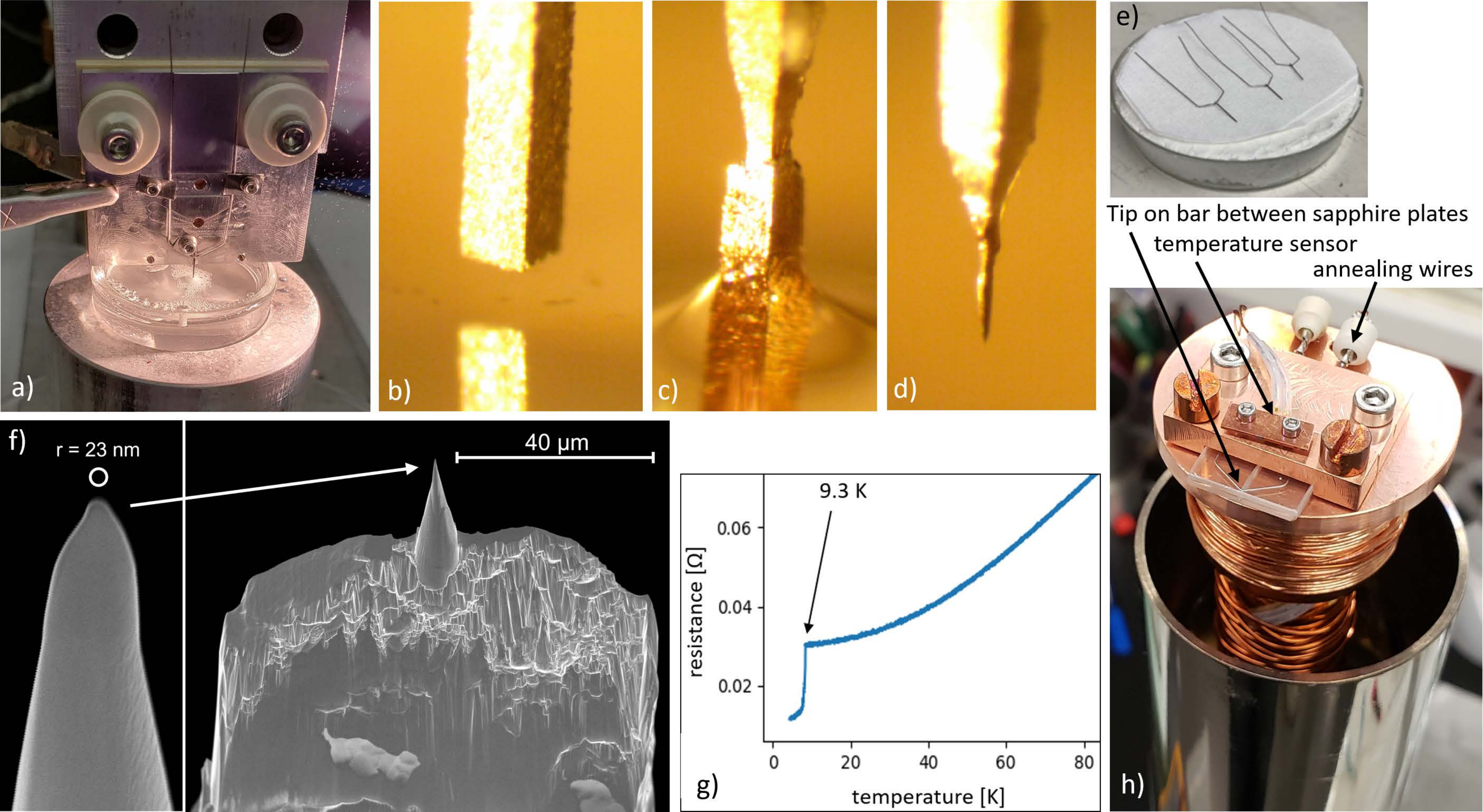} \caption{Tip fabrication: a) Setup for the electrochemical etching of the monocrystalline niobium tip that is spot-welded on a cathode holder and submerged in a KOH solution. b) Single crystal rectangular Nb wire before etching, c) during etching and d) after etching. e) Monocrystalline Nb wires spot-welded on a V-shaped bar before etching. f) Tip after ion milling in the FIB. Left side: magnification of the apex. Right side: Tip on shaft. g) 4-point measurement of the resistivity of the bar wire vs.~the tip temperature, with the superconducting transition step at \SI{9.3}{K}. h) The cryostat head is positioned upside down without the cooling shield indicating the tip mounting between two sapphire plates.}
\label{figure1}
\end{figure*}

In this article, such a novel source is described. We demonstrate the fabrication of a monocrystalline niobium (Nb) tip electron field emitter and analyze the emission properties at a superconducting temperature of \SI{5.2}{K}. The emitter has extremely low energy spreads, high beam currents, and different modes of operation that depend on the tip's apex surface geometry. This paper complements our recent work in \cite{Johnson2022}, where we demonstrated that a nano-protrusion (NP) can be formed on top of such a Nb tip by a specific annealing procedure, causing a distinct self-focusing field geometry. We will further label such an emitter as nano-protrusion tip (NPT). The space confinement in the NP leads to localized quantum band states at the apex \cite{Johnson2022,binh_field-emission_1992,purcell_field_1994,purcell_64_1995,vu_thien_binh_local_1992,gohda_total_2001}. The field emission energy spectra are Lorentzian-shaped and can be shifted in energy relative to the sharp, low-temperature Fermi edge, cutting-off the energy distribution for even smaller energy widths. This leads to energy distributions down to \SI{16}{meV} full width at half maximum (FWHM), an emission angle of 3.2$^\circ$, and a high reduced brightness of up to 5.0 $\times$ 10$^8$ A/(m$^2$ sr V) \cite{Johnson2022}. 

Here, we describe a different geometrical situation with a monocrystalline Nb field emitter with a radius of \SI{23}{nm} and without a NP on the apex. We will label it in the following as a Nb tip, in contrast to the NPT. The absents of the NP changes the associated geometry-dependent electronic band structure, leading to significant variations in the emission energy spectrum and the beam current behavior compared to a NPT. The first part of this article describes in detail the Nb tip and NPT fabrication by electrochemical etching and FIB (focused ion beam) ion milling, followed by annealing steps. Then we present our field emitter test and characterization setup and provide a theoretical analysis of the Nb tip and NPT field emission process. In our results section, we give examples of the energy spectra for a NPT and Nb tip and an intermediate state. The beam current at different Nb tip voltages is described, and we provide a beam stability analysis over several hours. The Nb tip emission tends to be less prone to adatom-related fluctuations than for a NPT and fits well the Fowler-Nordheim (FN) theory \cite{forbes_improved_2013,young_theoretical_1959}, as most commercial field emitters in microscopy do. However, at low temperatures, the measured field emission energy distribution is still significantly more monochromatic than conventional electron beam sources with around \SI{110}{meV} full-width-at-half-maximum (FWHM). This can be sufficiently narrow for several techniques in electron microscopy and spectroscopy. By determining the temperature dependence of the field emission spectra, we could demonstrate for a Nb tip and a NPT that the energy width changes only moderately between liquid helium and liquid nitrogen temperatures. This is of relevance for technical and commercial applications of our source. 

We furthermore demonstrate that a layer of xenon adatoms on the Nb tip can increase the field emission by two orders of magnitude while not significantly changing the energy distribution. Finally, we will discuss the role of superconductivity in the emission characteristics. There have been speculations \cite{oshima_monochromatic_1999} and theory predictions \cite{yuasa_entanglement_2009} that electrons connected as a Cooper pair inside the superconducting tip may get emitted into vacuum in a correlated or even entangled state with opposite momentum and spin. We measured the two-electron correlation with nanosecond resolution and could not confirm this phenomenon. However, we point out that this could be due to Nottingham heating \cite{vu_thien_binh_local_1992,xn_energy-exchange_nodate,bergeret_nottingham_1985,nottingham1941remarks} where the tip is locally heated due to the energy difference between the emitted electrons and the succeeding bulk electrons. It may cause the nanoscopic beam exit area on the tip apex to surpass the superconducting transition temperature. We discuss options to reverse this effect, leading to Nottingham cooling \cite{purcell_field_1994} and potentially realizing a correlated electron field emitter. Such an entangled electron source would have a significant impact on spectroscopy and quantum information science. Our observations are compared with previous field emission studies from superconducting tips \cite{gomer_field_1952,klein_field_1961,shkuratov_heating_1995,nagaoka_field_1999,nagaoka_monochromatic_1998,oshima_monochromatic_1999}. 

The Nb tip electron beam source presented in this article opens up new fields in high-resolution spectroscopy, limits the effect of aberrations, and will improve electron energy loss spectroscopy (EELS). It will decrease the impact of chromatic aberrations in low-voltage scanning electron microscopes, and when combined with a monochromator, it has the potential to enhance the energy resolution to the \SI{1}{meV} level.\\

\noindent {\bf \large Experimental setup}\\
\\
\noindent {\bf Tip preparation:} 
We fabricated the monocrystalline Nb tip shown in Fig.~\ref{figure1} e) in a four-step procedure. In case a NP should be generated on the apex, such as in \cite{Johnson2022}, a fifth step is added. In most of the literature on Nb field emitters, polycrystalline wires were used as a base for tip fabrication \cite{nagaoka_field_1999,nagaoka_monochromatic_1998,oshima_monochromatic_1999,hou_nanoemitter_2016}. Only two studies prepared a monocrystalline Nb source. One of them realized a [111] tip by electrochemical polishing from a single-crystal wire and measured the room temperature energy distribution \cite{nagaoka_energy_1996}. The other study did not prepare a tip, but analyzed the field emission current from crystalline Nb surfaces at high voltages at room temperature \cite{dangwal_pandey_field_2009}. For the generation of a beam with a narrow energy distribution, a monocrystalline structure is believed to be preferred, because the electrical resistance ratio at room temperature relative to \SI{4.2}{K} depends on the crystalline quality \cite{nagaoka_field_1999,takeuchi1980refinement}.

\begin{figure}[t]
\centering
\includegraphics[width=0.5\textwidth]{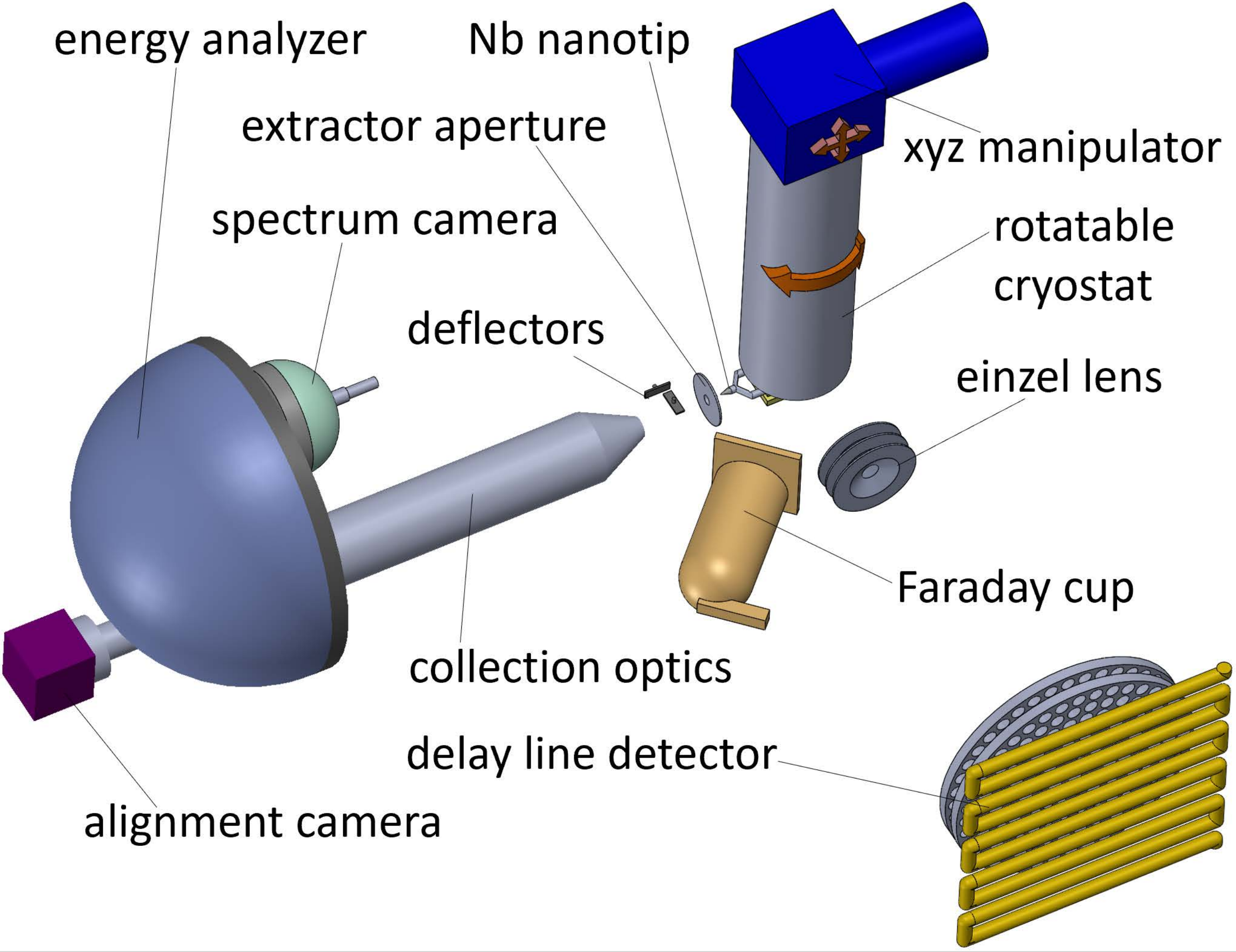} \caption{Sketch of the experimental setup for the in-vacuum characterization of the cryogenic niobium field emitter. It allows measuring the beam energy distribution, emission current, the angular profile and the electron-electron correlations. 
}
\label{figure2}
\end{figure}

Our tip preparation started with a larger Nb [100] oriented single crystal with a purity of 4N+ that was cut in rectangular monocrystalline pieces of 0.25$\times$0.25$\times$\SI{10}{\milli\metre}, as shown in Fig.~\ref{figure1} b), by EDM (electrical discharge machining, Surface Preparation Laboratory B.V). One of these pieces was then spot-welded on a polycrystalline Nb wire bar with the same diameter (\SI{0.25}{mm}) and bent to the V-shaped form as visible in the right inset of Fig.~\ref{figure1} e). In the next step, the monocrystalline piece was etched to a tip based on a procedure described in \cite{hou_nanoemitter_2016} and with the setup illustrated in Fig.~\ref{figure1} a). The monocrystalline wire on the bar was cleaned with isopropanol, acetone, and demineralized water in an ultrasonic bath. It was clamped on a holder that serves as an anode and immersed in a 5 molar potassium hydroxide (KOH) solution for electrochemical etching. A graphite electrode placed in the solution served as the cathode (not visible in Fig.~\ref{figure1} a). The immersion depth of the niobium wire should be $\sim$\SI{2}{mm}, which was set by a micrometer stage and monitored by an optical microscope using a camera. The etching process starts by applying an AC voltage, \SI{50}{Hz}, \SI{20}{Vpp}, between the cathode and anode. At the surface, a neck forms, narrowing the wire until the lower part (in the solution) falls off after about \SI{20}{min}. More KOH solution is constantly added  during etching to compensate for evaporation and to keep the neck at the liquid's surface. This creates the $\sim$\SI{80}{\micro \m} diameter tip shaft, visible in Fig.~\ref{figure1} e), which serves as a base for ion beam milling. For the fabrication of sharper tips, pulse waveform etching as described in \cite{hou_nanoemitter_2016} can be applied. Fig.~\ref{figure1} b), c) and d) show the tip before, during and after the etching process, respectively. Finally, the tip is cleaned by immersing it in isopropanol and deionized water and stored in vacuum or a dry gas atmosphere to avoid oxidation.

In the fourth step, the tip is then further shaped by gallium ion beam milling in a FIB (Focused Ion Beam, Thermo Fisher Scientific, model FEI Helios G4 UX) to cut the base symmetrically and form a conical Nb tip. The tip was imaged by SEM in the same instrument, as shown in two magnifications in Fig.~\ref{figure1} e). The left inset at high magnification indicates a tip radius of $\sim$\SI{23}{\nano\metre}. After this procedure, the tip was removed from the FIB and installed on a closed-cycle liquid helium cryostat (Advanced Research Systems, model DE-210). Fig.~\ref{figure1} g) illustrates that it is mounted between two sapphire plates, allowing electrical isolation and thermal conductivity. It is further thermally isolated by a copper cooling shield with a small aperture for the beam path, which was removed for the picture in Fig.~\ref{figure1} g). We confirmed superconducting conditions by cooling the tip to \SI{5.2}{K}, which is well below the transition temperature of Nb ($T_c =$ \SI{9.3}{K}), and performed a 4-point resistivity measurement while the cryostat's temperature increased. The expected step in resistivity at $T_c$ can clearly be observed in Fig.~\ref{figure1}~f) and was further used for temperature calibration. 

\begin{figure*}[t]
\begin{tikzpicture}
    \node[scale=1,inner sep=0pt] at (0,0) {\includegraphics[width=1\textwidth]{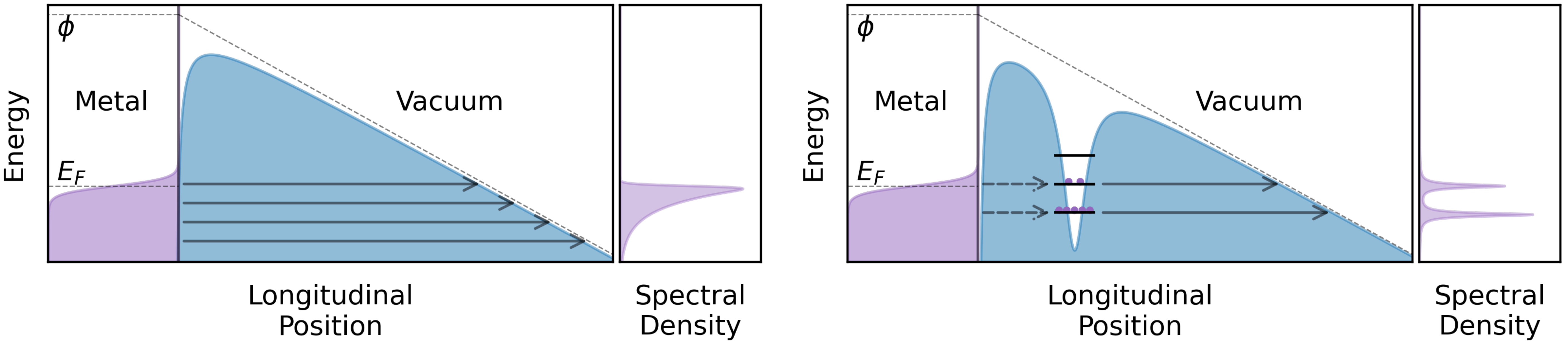}};
    \node[scale=1,inner sep=0pt] at (-8.8,1.8) {(a)};
    \node[scale=1,inner sep=0pt] at (0.35,1.8) {(b)};
    
\end{tikzpicture}
\caption{a) Illustration of the Fowler-Nordheim field emission process from a Nb tip that is set on a high voltage. The electrons tunnel through the Coulomb barrier from states around the Fermi-energy E$_F$ into vacuum. The inset on the right indicates the single broad energy peak in the spectra that is highly asymmetric due to the sharp cutoff at E$_F$ at low temperatures. b) The emission process from a NPT, where quantum band states form in the NP, acting as intermediate levels in the Coulomb barrier. The associated peaks in the field emission energy spectrum are Lorentzian distributed with significantly smaller widths. They can be shifted relative to E$_F$ by the applied tip voltage.  } 
\label{figure3}
\end{figure*}

A stable and intensive field emission requires stringent surface cleaning and ultrahigh vacuum conditions. In  \cite{nagaoka_field_1999,nagaoka_monochromatic_1998,oshima_monochromatic_1999} the Nb tip was prepared with repeated surface cleaning by field evaporation with a high voltage of +\SI{6}{kV}. Here, we clean the tip by annealing to $\sim$\SI{1220}{K} where the tip is clearly glowing. The annealing is done, while the cryostat is still on, by ramping up a current of \SI{4.85}{A} through the polycrystalline Nb wire bar. To avoid the tip getting blunt, an electrical bias of \SI{-3}{kV} is set on the extractor aperture during the process. The temperature is monitored through a vacuum window by a disappearing filament pyrometer, and the procedure is repeated several times. It is assumed that the gallium atoms from the FIB are removed at such high temperatures. The tip geometry can vary depending on the annealing duration, ramping speed, temperature, and number of heating cycles. This results in either a Nb tip with $\sim$\SI{23}{nm} radius or, as described in detail in \cite{Johnson2022,hou_nanoemitter_2016}, in the formation of a NPT with a nano-protrusion smaller than \SI{5}{nm} \cite{Johnson2022,binh_field-emission_1992} on the tip apex. We automated the annealing process with a programmable current source and could form the desired geometry in most of the cases. Sometimes, there are still differences in the outcome, altering the emission properties of the tip after a certain annealing cycle. This changes the onset voltage of the field emission, which was observed anywhere between $V_{tip}=$ \SI{-300}{V} and \SI{-500}{V}.\\

{\noindent \bf Field emitter characterization setup:}
The setup to measure the field emitter properties is illustrated in Fig.~\ref{figure2}. The vacuum pressure needs to be extremely low during the cleaning procedure \cite{oshima_monochromatic_1999} and field emission. Thus, the base pressure in our setup is kept at ~$\sim$\SI{8e-11}{Torr} by a large NEG pump in combination with a turbo pump. The chamber contains the closed-cycle liquid helium cryostat with a heater element for temperature control of the Nb tip between 5.2 and \SI{82}{K}. The cryostat is placed on a rotational flange and a 3D-manipulator, so the tip can be rotated and pointed towards three ports for measuring different beam features without breaking the vacuum. The first one is a hemispherical electron energy analyzer (ScientaOmicron DA20 R) with a resolution of \SI{3} {meV}. For field extraction and beam guiding, an extractor aperture and a custom deflector element are installed between the tip and the analyzer's entrance and collection optics. A cryostat rotation by 45$^\circ$ points the tip towards a Faraday cup (Kimball Physics, FC-71) with a pico-amperemeter (Keithley, Model 237). It allows recording Fowler-Nordheim plots of the beam current as a function of the applied tip voltage. Additionally, it can perform long-term emission stability measurements. After a 90$^\circ$ rotation, the emitter can be pointed towards a single electron delay line detector (RoentDek DLD HEX100) with a high spatial and temporal resolution. It allows imaging of the beam profile with the angular distribution at low intensity. Along the beam path, a custom Einzel-lens for beam magnification is positioned, followed by a deflector for alignment. It spreads the emission to a large area on the MCP, which is necessary to determine spatial and temporal electron-electron correlations on the nanosecond scale.\\

\begin{figure}[t]
\centering
\includegraphics[width=0.48\textwidth]{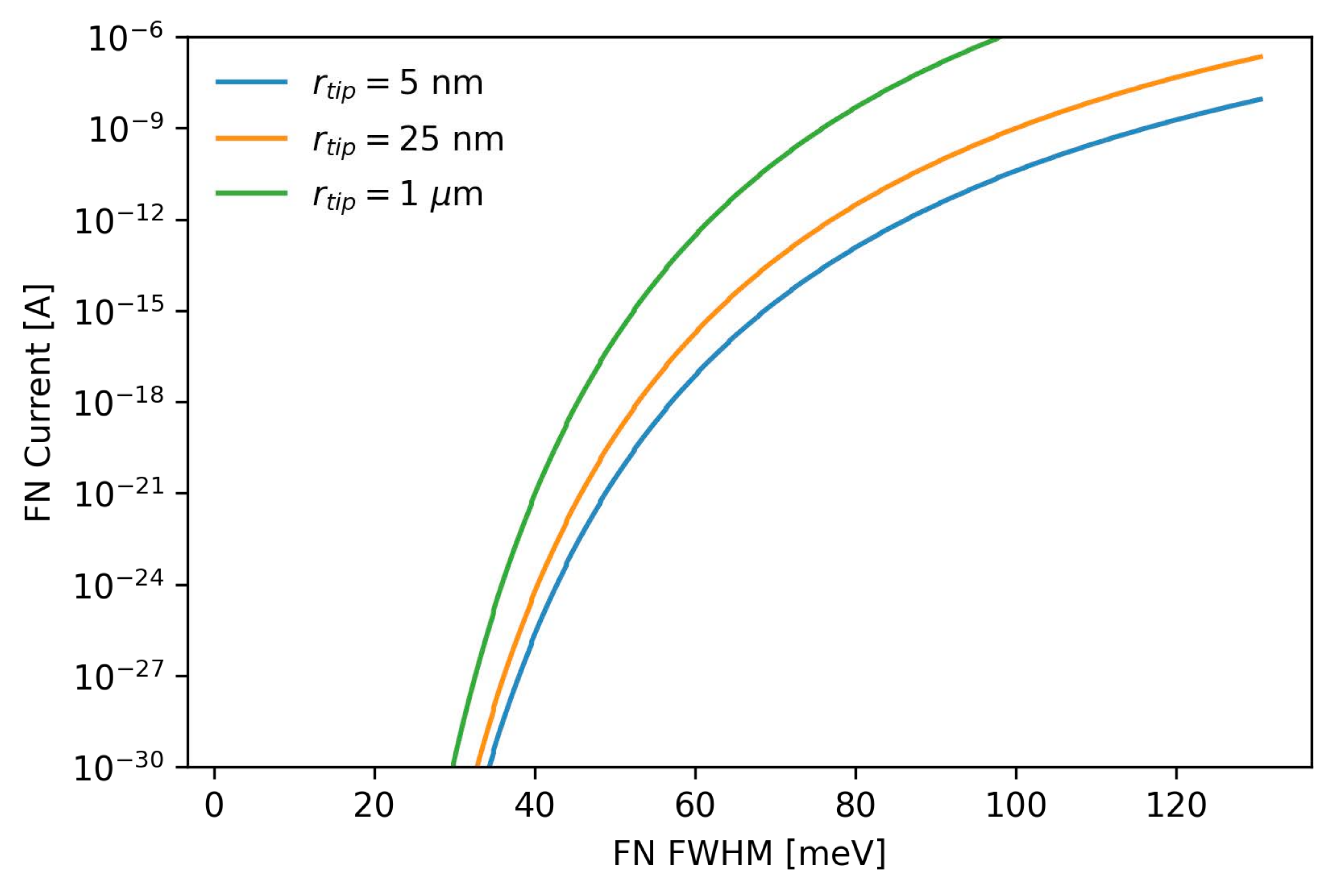} \caption{The dependency of the electron beam current as a function of the energy distribution FWHM for three standard field emitters that can be described by the Fowler-Nordheim relation with different radii of their emitting areas at $T=0$ K. Reasonable currents for microscopy applications in the nA-regime lead to energy widths around \SI{100}{meV}. }
\label{figure4}
\end{figure}

\noindent {\bf \large THEORY} \\
Electrons in a metallic cathode are kept from escaping to vacuum by the work function $\phi$, an electrostatic barrier above the Fermi level $E_F$. This potential barrier can be decreased by a negative bias voltage applied to the cathode or a positive bias on the extraction aperture. It further gets decreased by the image charge potential due to the interaction of the beam electrons with the conduction electrons in the metal:
\begin{equation}
    V_{pot}(x) = \phi-eFx-\frac{e^2}{16\pi\epsilon_0x} \label{eq:potential}
\end{equation}
where $F=|\bE|=\beta V_{tip}$ is the magnitude of the electric field at the emission region, with $\beta\approx$ \SI{0.008}{\nm\tothe{-1}} being an enhancement factor that is a function of the tip's geometry at an applied tip voltage $V_{tip}$ \cite{biswas_tunneling_2018}.
The total energy distribution, with respect to $E_F$, of electrons emitted from a finite temperature metallic tip cathode in the Murphy-Good regime between pure field emission and thermal emission can be written as 
\begin{equation}
    G(E,F,T,\phi) = \frac{4\pi m}{h^3}\frac{f(E,T)D(F,\phi)\exp(E/D(F,\phi))}{\exp(B(F,\phi)\phi^{3/2}/F)}, \label{eq:FN_E}
\end{equation}
where $T$ is temperature, $f(E,T)=(1+\exp(E/k_BT))^{-1}$ is the Fermi-Dirac distribution, $B(F,\phi)=8\pi\sqrt{2m}v(y)/3h$ and $D(F,\phi) = e\hbar F/2\sqrt{2m\phi}$ with $v(y)\approx1-y^2(3-\ln(y))/3$ and $y=\sqrt{e^3F/4\pi\epsilon_0\phi^2}$ \cite{young_theoretical_1959}. The tunnel barrier increases linearly in ($E_F - E$) with $E<< \phi$, resulting in an exponentially decaying emission probability below $E_F$. There is a hard cut-off in the emission probability above $E_F$ due to the sharp exponential tail of the Fermi-Dirac distribution at low temperatures. This is illustrated in Fig.~\ref{figure3} a). Integration over $G$ and multiplying by the electron charge $e$ gives the well-known FN current density, $J(F,T,\phi) = e\int dE\, G(E,F,T,\phi)$ \cite{murphy_thermionic_1956}. As shown in Fig.~\ref{figure4}, the width of the FN energy distribution from standard emitters is limited. Depending on the emitting area (blue, orange and green lines in Fig.~\ref{figure4} for emission radii of r$_{tip}=$ 5, 25, \SI{1000}{nm}) the FWHM energy distribution for nA beam currents that are usually required for electron microscopy, are in the range of \SI{100}{meV} at $T=0$ K.

However, for a significantly smaller field emission energy distribution at a comparable beam current, a different emission process needs to be established that leverages distinct quantum states due to the surface geometry of extremely sharp tips. Equation (\ref{eq:FN_E}) can be used to model the total energy distribution and current density from tip emitters with tip radii down to a single-atom tip. But if the height to width ratio of the tip exceeds some critical threshold, such as in the NPT, the confinement potential creates localized discrete electronic states away from the bulk Fermi sea. These states are commonly approximated as a well in the Coulomb barrier, spatially separated from, yet supplied directly by, the Fermi sea \cite{binh_field-emission_1992}, as illustrated in Fig.~\ref{figure3}~b). The discrete states of the NP act as an intermediary between the Fermi sea and vacuum, giving a resonant enhancement of tunneling probability at the energy of the mid-barrier state; a process resembling the physics of single-atom resonance-tunneling spectroscopy \cite{gadzuk_resonance-tunneling_1970}. If we neglect the interference between the direct and resonant tunneling amplitudes, the NPT emission total energy distribution can be written as
\begin{equation}
    G_{tot}(E,F,T,\phi) \approx G(E,F,T,\phi)\left(1+\sum_nR_n(E,F)\right). \label{eq:FN_E_RT}
\end{equation}
where the resonant enhancements factors $R_n$, have the form
\begin{equation}
     R_n(E,F) = \frac{A_n}{(E-E_n+\alpha F)^2+\Gamma_n^2}, \label{eq:FN_E_RTn}
\end{equation}
i.e., have a Lorentzian shape centered at the mid-barrier state energy $E_n$ that shifts in energy proportionally with the applied field, an associated linewidth of $\Gamma_n$, and a resonance magnitude of $A_n$ \cite{gadzuk_resonance-tunneling_1970}. They are single-peaked functions corresponding to each discrete state of the NP at an energy that shifts linearly with the electric field strength at the tip. 
With increasing field strength $F$, the Coulomb barrier to vacuum is lowered, shifting the NPT field emission peaks down in energy with respect to $E_F$. As these resonances are multiplied by the envelope of FN emission, resonant tunneling is most efficient through states near $E_F$ and the applied field required for resonant tunneling can be much lower than what is typically necessary for a tip without a NP. Consequently, resonant emission from a discrete NPT state can be present without the FN emission with the broad energy spectrum and NPT emission can be realized with linewidths in the order of tens of \SI{}{\milli\eV} \cite{Johnson2022,purcell_64_1995}. Extremely narrow field emission energy spectra peaks can be achieved when the tip is cold enough that the width of the Fermi-Dirac distribution is narrower than the linewidth of the discrete state and the center energy is near $E_F$. Then the peak shift with the applied field can be used to adjust the emission peak such that the Fermi-edge of the distribution cuts off a portion of the peak, lowering the linewidth of the total energy distribution of the emission \cite{Johnson2022,purcell_field_1994}. We demonstrated this process with the described setup in \cite{Johnson2022}.
The physics of this observation is similar to an early method in atomic spectroscopy for atoms adsorbed to metallic surfaces \cite{gadzuk_resonance-tunneling_1970}, and similar resonant field emission spectra have been demonstrated with gold nanoclusters deposited on tips \cite{lin_observation_1991}, single-atom nano-protrusions grown in-situ on tips by high field and heat \cite{binh_electron_1992}, edge states from graphene on tips \cite{diehl_narrow_2020}, and photo-assisted emission from a quantum dot on a tip \cite{duchet_femtosecond_2021}. It has been shown that the exact energy distribution of emission from nano-protrusions can be determined by density functional calculation \cite{gohda_total_2001}. However, there is a strong dependence on the exact geometry of the apex and uncertainty in the growth process \cite{binh_electron_1992}. Thus, it can be more useful to take an empirical approach and assume a resonant enhancement factor that is similar to resonant tunneling through an adsorbed surface atom \cite{purcell_field_1994}. This has been derived rigorously for a deposited gold cluster in \cite{lin_field-emission_1992}. \\

\begin{figure}[t]
\centering
\includegraphics[width=0.48\textwidth]{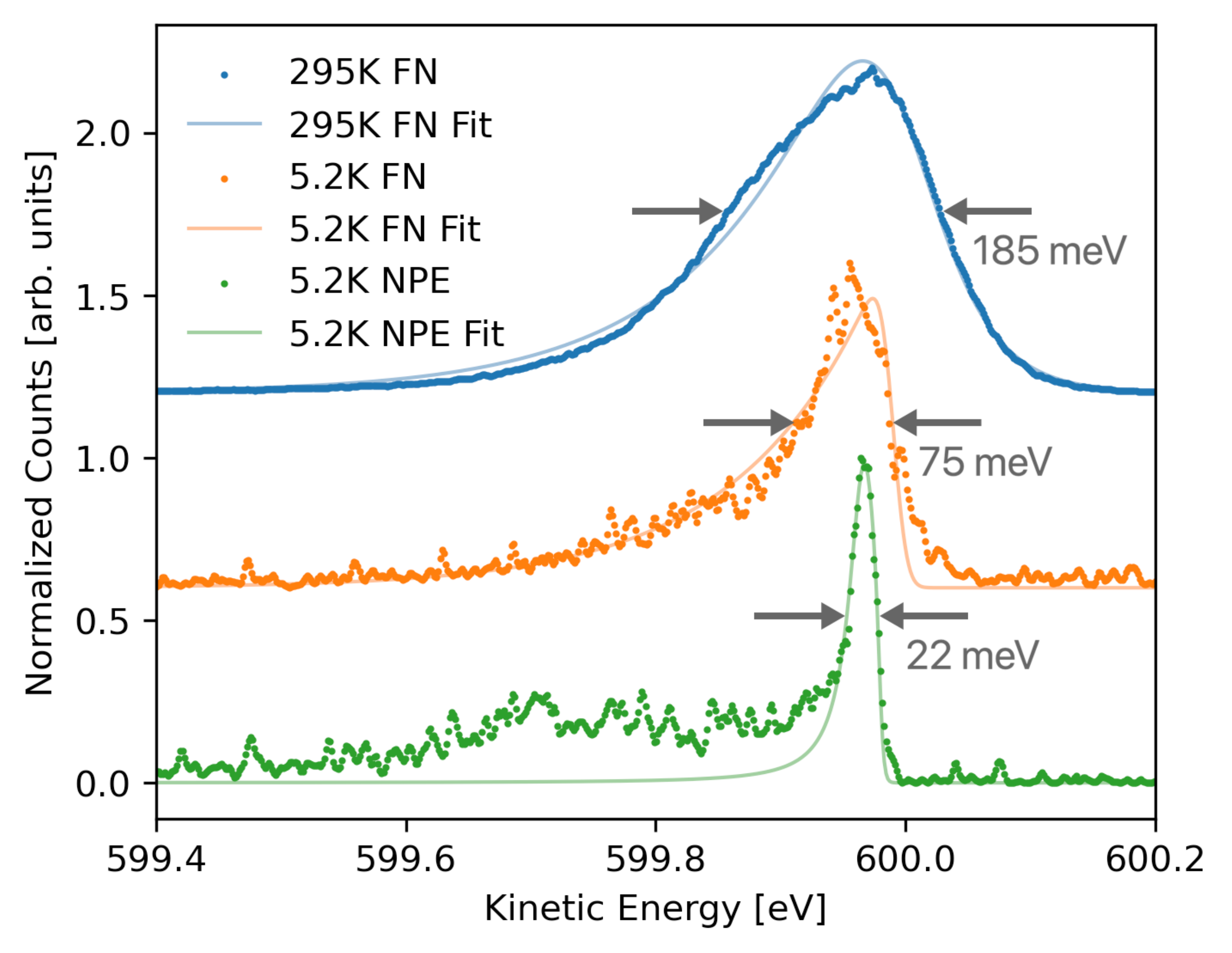} \caption{Comparison of the Nb tip field emission energy distribution at the Fermi edge without a NP at room temperature (blue upper curve), at a superconducting temperature of \SI{5.2}{K} (orange middle curve), and after the formation of a NP on the tip apex also at \SI{5.2}{K} (green lower curve).}.
\label{figure5}
\end{figure}

\noindent {\bf \large RESULTS} \\
We describe the field emission of the Nb tip shown in Fig.~\ref{figure1}~f) without a NP in a superconducting state at \SI{5.2}{K} and compare it with a NPT. The upper blue curve in Fig.~\ref{figure5} presents the energy spectrum of the Nb tip (without NP) field emission at room temperature and at a tip voltage of $V_{tip} =$ \SI{-600}{V}, with a FWHM of \SI{185}{meV}. As expected from the FN theory, it is significantly broader than the emission from the same tip at \SI{5.2}{K}, with a FWHM of \SI{75}{meV} (orange middle curve). After a NP was formed on the tip's apex by the annealing procedure described in the preparation section, the NPT was cooled again to \SI{5.2}{K} and yielded an energy distribution with an ultra-narrow FWHM of \SI{22}{meV} (green lower curve), also at $V_{tip} =$ \SI{-600}{V}. As expected, the peak is Lorentzian-shaped and cut off by the sharp low temperature Fermi edge. Our measurements indicate clearly that the temperature reduction decreases significantly the energy width of the FN-emission of the Nb tip. Further decrease is feasible by changing the tip apex geometry by adding the NP, as described in the theory section. The field emission energy spectrum, brightness and current stability of a NPT are discussed in detail in \cite{Johnson2022}. There, a different monocrystalline Nb tip was applied, but both emitters were fabricated according to the same procedure as described above.

\begin{figure}[b]
\centering
\includegraphics[width=0.495\textwidth]{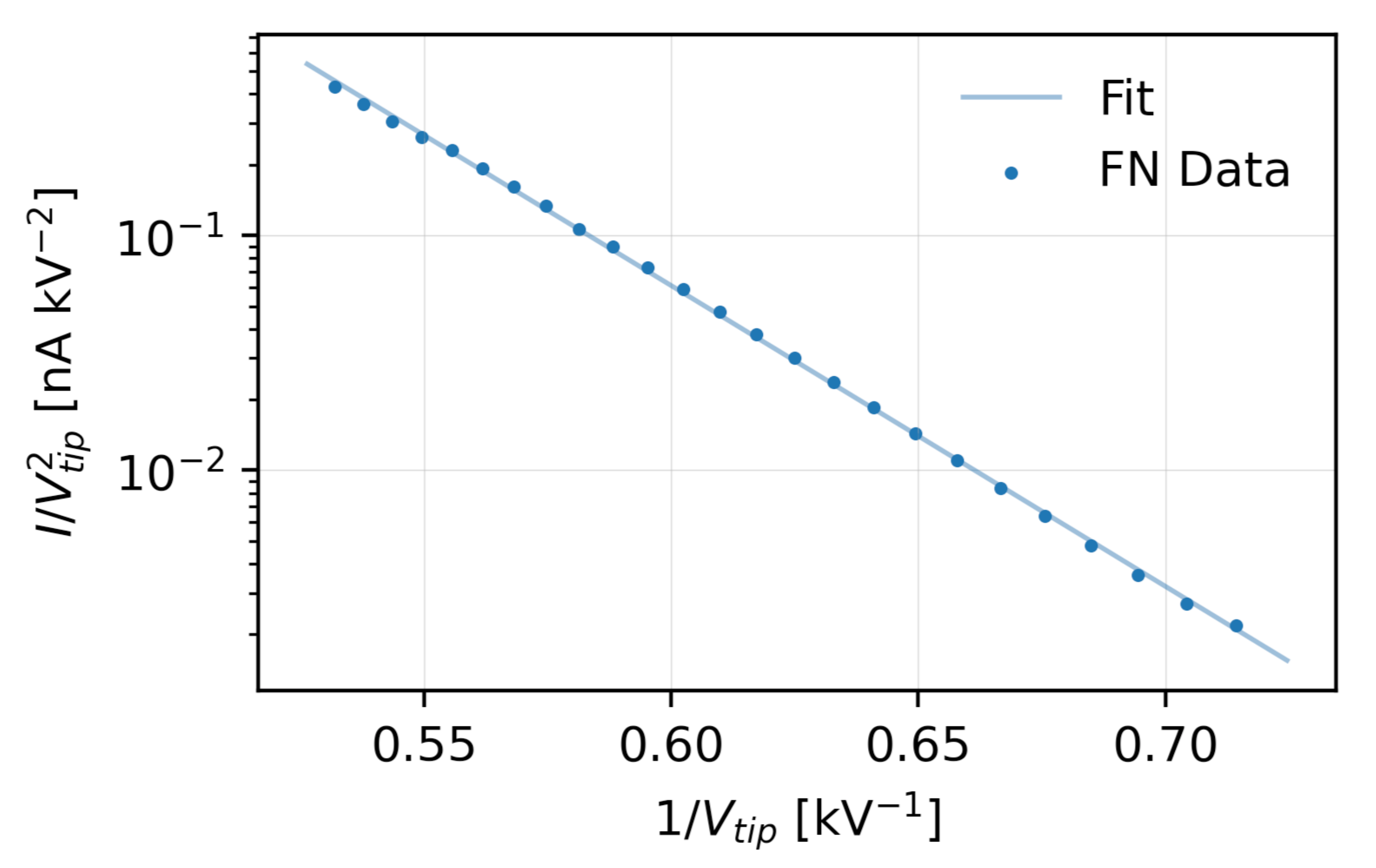} \caption{FN measurement with the Faraday cup of the field emission current vs.~the applied tip voltage for the superconducting, monocrystalline Nb tip shown in Fig.~\ref{figure1}~e). The data matches well with a linear FN fit function. }
\label{figure6}
\end{figure}

\begin{figure}[t]
\centering
\includegraphics[width=0.495\textwidth]{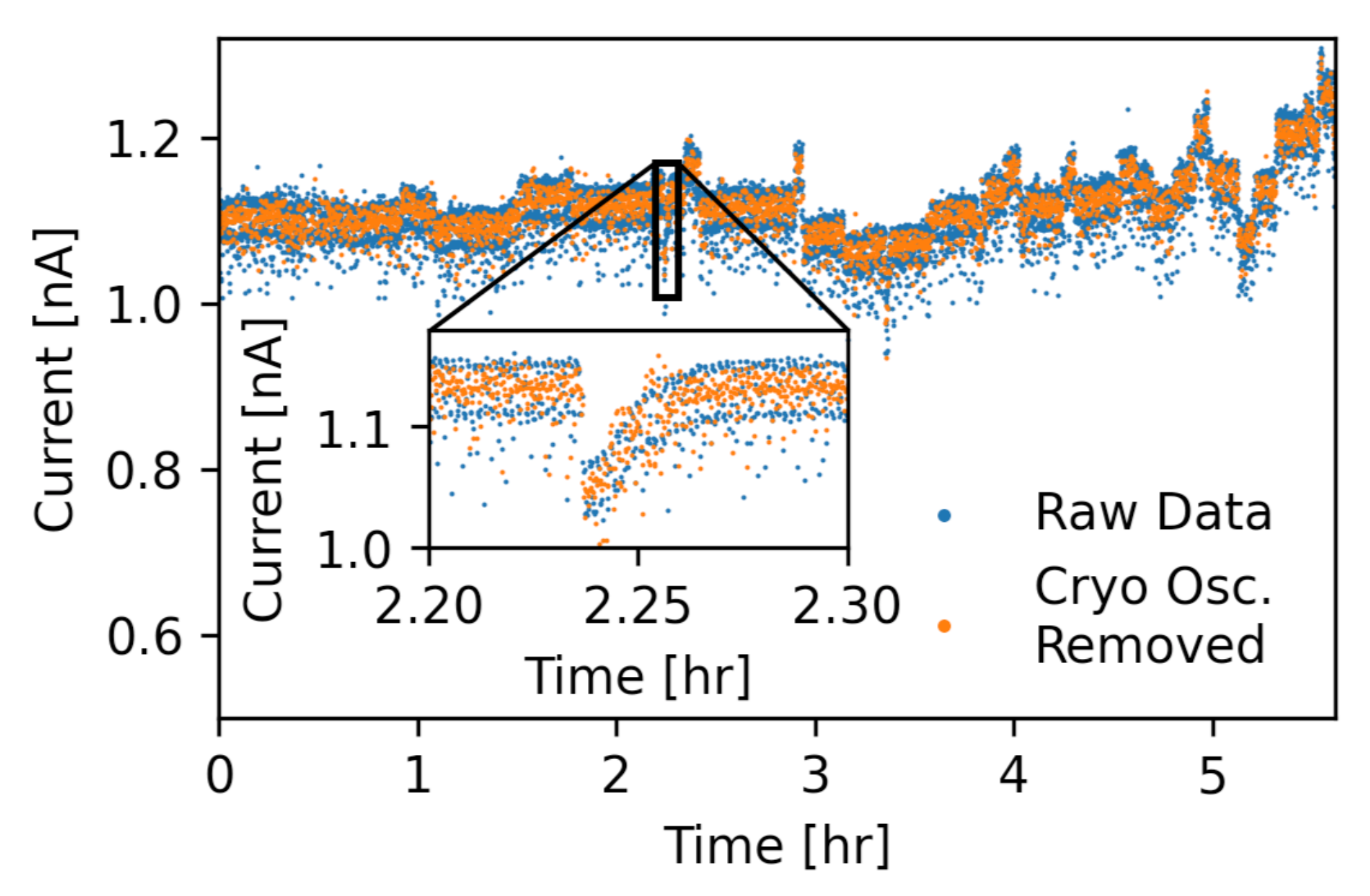} \caption{Stability measurement of the superconducting Nb tip field emitter beam current (blue points). The orange points represent the data after a Fourier-filter to remove the periodic current fluctuations from the closed-cycle cryostat pumping vibrations.}
\label{figure7}
\end{figure}

For further beam analysis of the superconducting Nb tip without the NP at \SI{5.2}{K}, we performed a FN current vs.~tip voltage measurement with the Faraday cup indicated in Fig.~\ref{figure2}. As it was already observed in the energy spectra of Fig.~\ref{figure5}~c), field emission from metal tips follows the described FN current density $J(F,T,\phi)$. A Faraday cup measurement is expected to give a linear relationship between the natural log of emitted current divided by tip voltage squared as a function of inverse tip voltage \cite{forbes_improved_2013}. This is perfectly reflected in our Faraday measurement shown in Fig.~\ref{figure6} for a Nb tip field emission at a temperature of \SI{5.2}{K}, where the data points can be well-matched by a FN-fit function. The observed emission is in clear contrast to such a measurement for the Nb tip with a nano-protrusion at the apex in \cite{Johnson2022}. There we determined a clear deviation from the FN theory for the ultra-narrow discrete state emission. 

Furthermore, we performed a long-time field emission current stability measurement at \SI{5.2}{K} monitoring the beam current over 5 hours with the Faraday cup for the Nb tip (without the nano-protrusion). The data is shown in Fig.~\ref{figure7}. The closed-cycle cryostat pumps liquid helium with a period of 1.5 seconds, which causes the tip to be mechanically displaced by about 50-\SI{100}{\micro m} in every pump cycle relative to the fixed extraction aperture. This movement changes the field configuration throughout the pump cycle causing slight deviations in the field emission current as can be observed in the blue measurement points of Fig.~\ref{figure7}. We did a Fourier analysis of the data for an emission current of around \SI{1.1}{nA} and removed the fluctuations due to the periodic cryostat's vibration frequency, leading to the orange dots in Fig.~\ref{figure7}. For the long-time stability, drifts and jumps of the cryo-tip beam current that are also known from cold field emitter made of other materials, play a more important role. This can also be observed in Fig.~\ref{figure7} in particular after \SI{3}{h} when adatom accumulation on the surface increases. We measured long-time fluctuations on the order of 10 \% of the mean tip current. \\

\begin{figure}[b]
\begin{tikzpicture}
    \node[scale=1,inner sep=0pt] at (-7,0) {\includegraphics[width=0.48\textwidth]{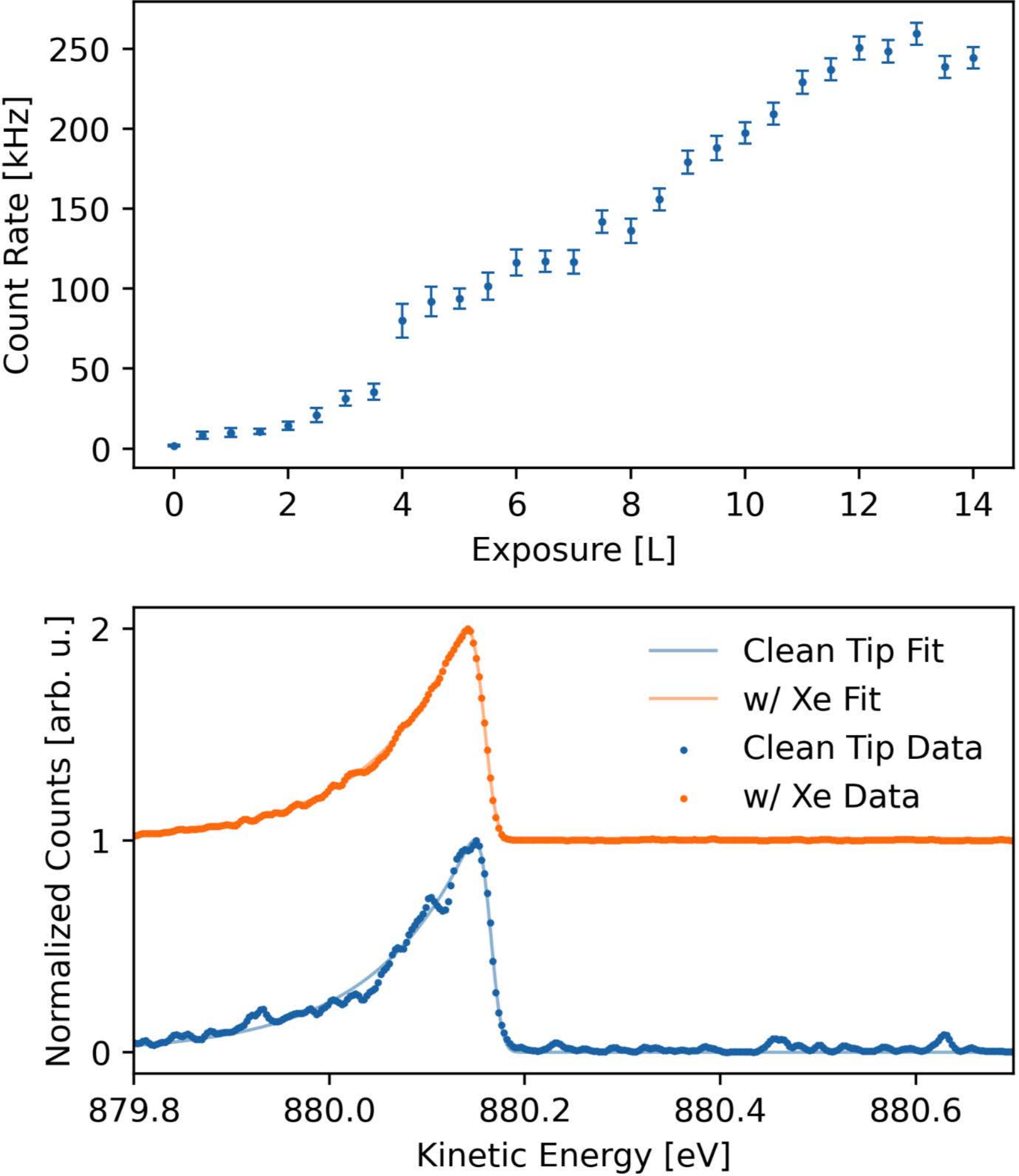}};
    \node[scale=1,inner sep=0pt] at (-11.14,4.85) {(a)};
    \node[scale=1,inner sep=0pt] at (-11.14,0) {(b)};
    
\end{tikzpicture}
 \caption{a) Increasing beam current with increasing coverage of xenon gas adsorbed on the superconducting Nb tip at a constant field emission voltage. b) Energy spectra with xenon coverage of \SI{14}{L} on the Nb tip (top orange curve) and for the clean Nb tip (lower blue curve), both with a FN fit.
}
\label{figure8}
\end{figure}

Introducing xenon (Xe) gas to the vacuum chamber with Nb tip field emitters at cryogenic temperatures has been shown to increase the emission current significantly \cite{hou_nanoemitter_2016}. This is due to the adsorption of the Xe atoms onto the tip surface, which is known to lower the effective work function \cite{hou_nanoemitter_2016}. 
We made similar studies with our superconducting emitters and set a sequence of Xe gas exposures to the Nb tip at a temperature of \SI{5.6}{K}. Every step increased the Xe adsorption by approximately \SI{0.5}{L} (1 Langmuir (L) = 10$^{-6}$ Torr s). Then we determined the field emission total count rate of the magnified angular distribution with the delay line detector at a constant tip voltage of \SI{-1150}{V}. The data is shown in Fig.~\ref{figure8}~a) revealing an increase in the count rate by a factor of $\times$116 at a Xe exposure of \SI{14}{L}. While field strength and tip geometry have long been used as parameters to tune the emission properties of field emitters, working with a cryostat for temperature control and adsorbed gases for work function lowering significantly expands the parameter space for tunable emitters. We also analyzed if the Xe-coverage changes the beam's energy distribution. As demonstrated in Fig.~\ref{figure8}~b), the field emission energy FWHM of the clean tip was \SI{69}{meV}. After \SI{14}{L} Xe-gas coverage, the FWHM was \SI{66}{meV}. The
difference is within our measurement error and both energy peaks are distributed according to FN, so we do not observe any change in the beam energy distribution after Xe-atom adsorption on the tip.

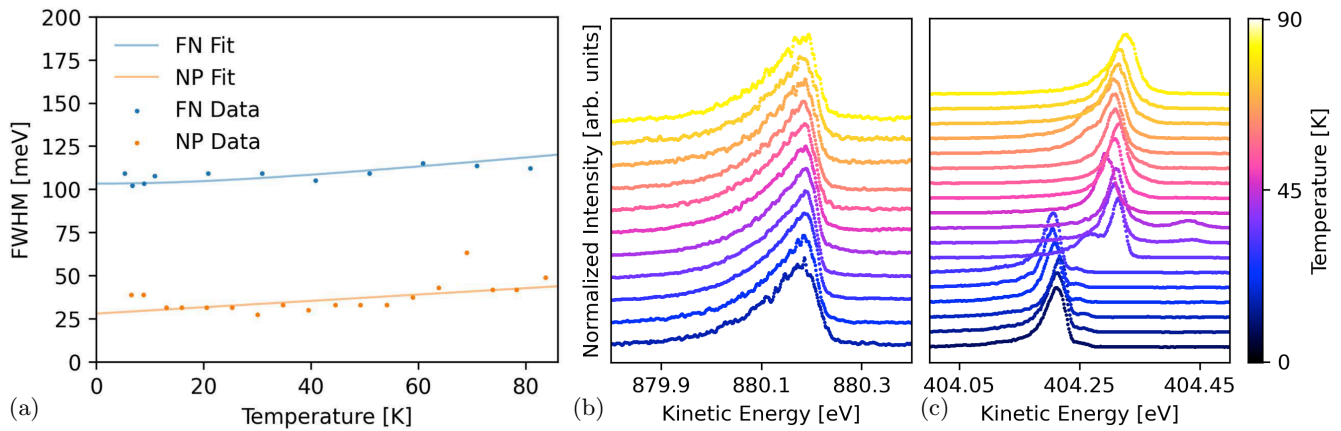
\begin{figure*}[t]
\begin{tikzpicture}
    \node[scale=1,inner sep=0pt] at (-3,0) {\includegraphics[width=1.0\textwidth]{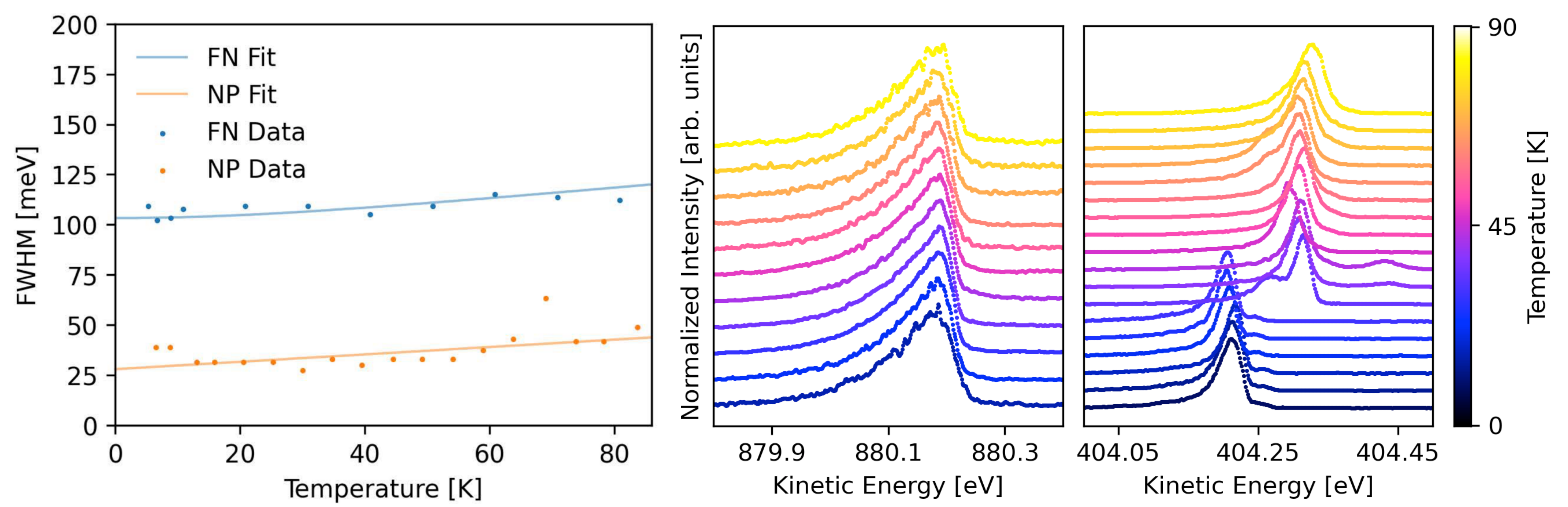}};
    \node[scale=1,inner sep=0pt] at (-11.6,-2.6) {(a)};
    \node[scale=1,inner sep=0pt] at (-4.1,-2.6) {(b)};
    \node[scale=1,inner sep=0pt] at (0.5,-2.6) {(c)};
    
\end{tikzpicture}
 \caption{a) Temperature dependence of the electron energy distribution FWHM in the beam from a Nb tip field emitter with (orange dots) and without (blue dots) a nano-protrusion on the apex. The blue line is a fit according to the FN theory including instrument noise and the orange one is a basic linear fit. b) Temperature-dependent FN emission spectra corresponding to the blue dots in a). c) Temperature-dependent nano-protrusion emission spectra corresponding to the orange dots in a). The origin of the sudden shift at $\sim$\SI{25}{K} by $\sim$\SI{10}{meV} is unknown.
}
\label{figure9}
\end{figure*}

As a next step, the temperature dependence of the field emission energy spectrum of our electron source was measured. Previous studies determined the beam energy width of cold platinum tip field emission from localized surface band states from \SI{80}{K} to \SI{293}{K} \cite{purcell_64_1995}. They demonstrated an energy distribution of \SI{64}{meV} at \SI{80}{K} and found a linear increase in the linewidth up to \SI{100}{meV} at \SI{293}{K}. Considering their calculated instrument broadening, it was predicted that the energy distribution could be decreased down to around \SI{20}{meV} by lowering the temperature close to \SI{0}{K} \cite{purcell_64_1995}. In \cite{Johnson2022} and Fig.~\ref{figure5} a) of this work, we could demonstrate this predicted ultra-narrow energy width with the Nb nano emitters with a nano-protrusion. Here, we compare the temperature dependence between \SI{5.6}{K} and \SI{82}{K} in Fig.~\ref{figure9} for the two described tip geometry configurations. The blue dots and the blue FN fit in Fig.~\ref{figure9} a) illustrate the measured temperature dependence of the FWHM of the beam's energy distribution for the Nb tip that follows the Fowler-Nordheim description. The orange dots with a linear fit exhibit the temperature dependence FWHM data of the resonant localized band state emission for the NPT (fabricated in \cite{Johnson2022}). The results reveal that there is only a moderate broadening in the linewidth FWHM from liquid helium (\SI{4.2}{K}) to liquid nitrogen (\SI{77}{K}) temperatures. This observation is quite significant for the integration of this field emitter in electron microscopes since it is technically much easier and cheaper to cool the emitter with liquid nitrogen. The outcome demonstrates that narrow energy distribution emission well below \SI{40}{meV} is feasible at liquid nitrogen cooling. The data also indicates a difference in energy width of $\sim$\SI{75}{meV} between the two tip geometries.

Another important question to address is the possible correlations of electrons emitted from the superconducting tip. According to a theory paper by Yuasa et al.~\cite{yuasa_entanglement_2009}, a niobium tip is a possible source of entangled free electrons with opposite spin and momentum following field emission of correlated Cooper pairs. However, it is still an open experimental question if a correlated two-electron emission can be realized from solid-state surfaces to the vacuum. Solid-state Cooper pair tunneling through a barrier is well known in physics, e.g.~it is the basis for superconducting quantum interference devices (SQUIDs) and is applied in superconducting scanning-tunneling experiments \cite{rodrigo_josephson_2006}. Also, a recent solid-state experiment has demonstrated efficient Cooper pair splitting into normal conductors in nanostructured devices \cite{hofstetter2009cooper}, resulting in spatially separated pairs of entangled electrons. A further theoretical study considered a superconducting tip in a high electric field \cite{chen_vortices_2008-1} and predicted the existence of vortices and superconducting states that persists in nano-sized tips up to fields much larger than the bulk critical field. They conclude that a significant Cooper pair density is present even in the confined space of a tip with a size on the order of the coherence length. Confirming entangled electron emission from a superconducting source would be a breakthrough in the field. Such an emitter would create significant opportunities for quantum information science with free electrons, in analogy to the scientific impact of entangled two-photon sources \cite{kwiat_new_1995}. They led to major achievements in quantum optics, including the violation of Bell inequality \cite{ou1988violation}, quantum cryptography \cite{jennewein2000quantum}, quantum teleportation \cite{bouwmeester1997experimental} and quantum computing \cite{walther2005experimental}. It would additionally allow correlated two-electron spectroscopy, with an unprecedented accuracy that depends on the extremely low energy difference between the correlated electrons rather than the single electron-electron energy variations in an uncorrelated beam.

\begin{figure}[t]
\centering
\includegraphics[width=0.5\textwidth]{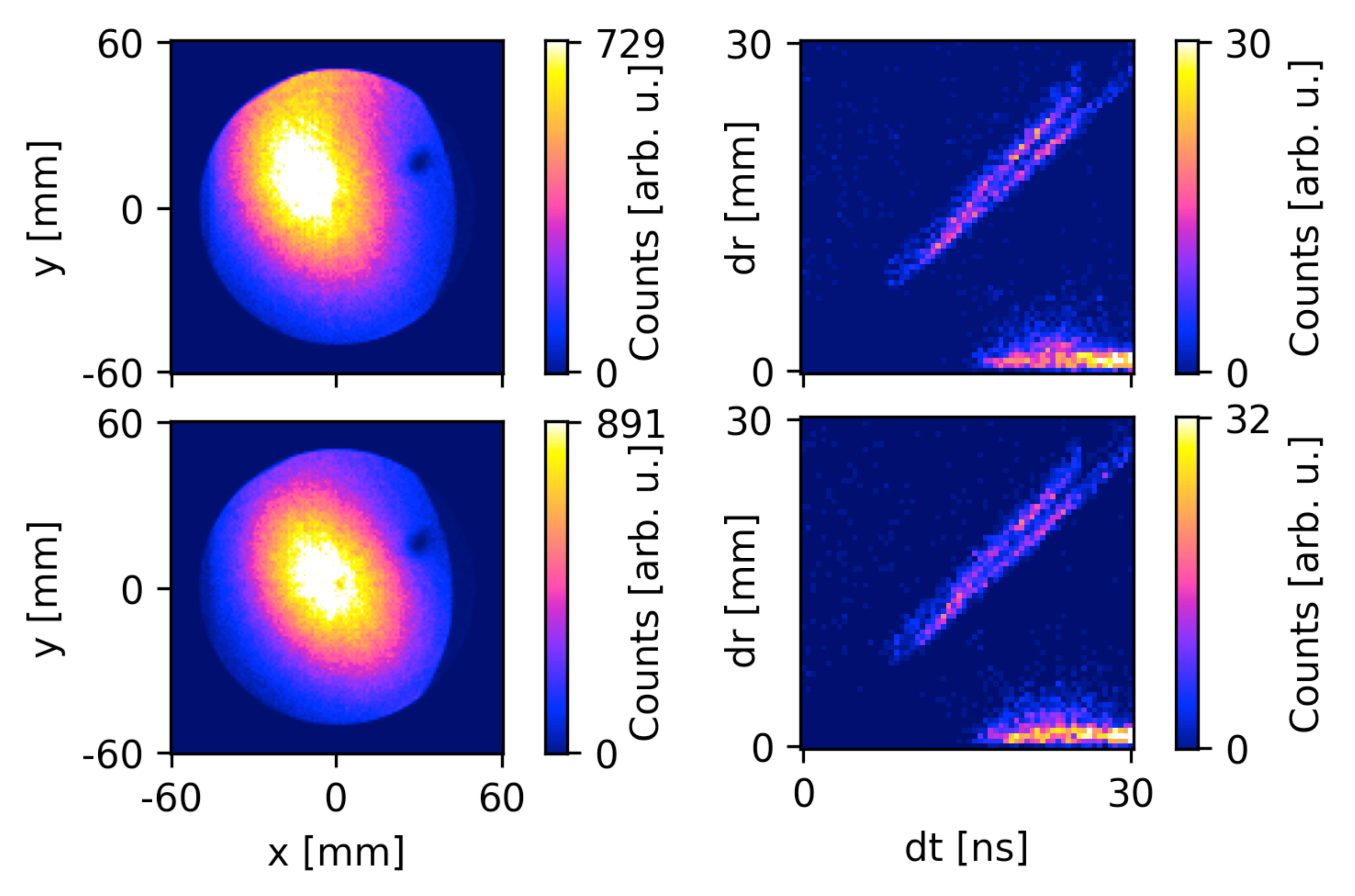} \caption{a) Integrated total electron beam signal on the delay line detector at a superconducting tip temperature of \SI{5.6}{K}. b)~Correlation analysis between two consecutive electrons with the time difference $dt$ and the spatial separation $dr$ at a superconducting tip temperature corresponding to panel a). c) and d) are integrated signal and correlated counts at a normal conducting temperature of \SI{44}{K}.
}
\label{figure10}
\end{figure}

For that reason, we performed an electron correlation analysis in the field emission of our Nb emitter. The delay line detector in our setup (see Fig.~\ref{figure2}) is capable to verify electron pair emission with high spatial and temporal resolution if the electrons are spatially separated by more than $\sim$\SI{8}{mm} \cite{jagutzki_broad-application_2002}. Based on the predicted opposite initial momentum of the correlated electrons \cite{yuasa_entanglement_2009}, they are expected to arrive at the detector with a distinct distance $dr$ within a few nanoseconds $dt$. To increase the local separation, the beam gets magnified by the einzel lens across the delay line detector area with a radius of \SI{50}{mm}. The results are shown in Fig.~\ref{figure10}~a) and b) for the superconducting tip temperature of \SI{5.6}{K}, and in Fig.~\ref{figure10}~c) and d) for a normal conducting tip temperature of \SI{44}{K}. In both cases, we integrate over $7\times 10^5$ counts. Fig.~\ref{figure10} a) and c) are the magnified angular beam profiles at normal and superconducting temperatures as they arrive at the detector, respectively. In Fig.~\ref{figure10} b) and d) we add the arrival time information to plot the time difference $dt$ between any two detected electrons up to \SI{30}{ns} against their spatial distance on the detector $dr$. It provides the correlated events under superconducting (Fig.~\ref{figure10} b) and normal conducting (Fig.~\ref{figure10} d) conditions. There is not a significant difference between the two measurements. The counts in the lower right corner with $dr \sim$\SI{0}{mm} and $dt \sim$20-\SI{30}{ns} are believed to be due to ion feedback in the detector, where a residual gas atom in an MCP channel gets ionized by an electron avalanche and accelerated in the opposite direction. Within the same channel, it triggers a second, delayed electron pulse. It is a known effect for MCP detectors. The origin of the diagonal features in Fig.~\ref{figure10} b) and d) is not clear, we assume double counts due to ringing effects in the detector electronics. However, these are only a few seemingly correlated counts in both images relative to the total accumulated signal. They are extremely rare, and their number and pattern are about the same at normal and superconducting temperature. For that reason, we do not observe any significant two-electron correlation in the superconducting field emission. \\

\noindent {\bf \large Discussion}\\
\\
We discussed the field emission properties of monocrystalline niobium tip electron beam sources at a temperature of \SI{5.2}{K} well below the superconducting transition. This work complements our recent publication \cite{Johnson2022} where we emphasized the resonant emission through band states with ultra-low energy distribution from a Nb nanotip with a nano-protrusion on the apex. Here, we concentrate on the emission behavior of such an Nb tip without the nano-protrusion and compare the two cases with each other. We provide a detailed description of the fabrication steps to produce a tip with a \SI{23}{nm} radius. The tip is cooled by a liquid helium cryostat. 
Energy distributions around \SI{100}{meV} are measured, with an emission current stability of \SI{10}{\%} in the nano ampere regime. 
The current increases with tip voltage according to the Fowler-Nordheim description. Additionally, it is demonstrated that the adsorption of xenon atoms leads to an increase in beam intensity by two orders of magnitude. 

We also compared the temperature behavior of the field emission for both emission processes from 5.6 to $\sim$\SI{82}{K}. It is observed that the energy width of the beam does not strongly change between liquid helium and liquid nitrogen cooling temperatures. With the NPT, narrow energy widths below \SI{40}{meV} were observed, and with the Nb tip they remain below \SI{110}{meV}. The cooling with liquid nitrogen eases the commercial application of this source compared to liquid helium significantly. According to an evaluation based on BCS theory by Gadzuk \cite{gadzuk_many-body_1969,nagaoka_field_1999}, a narrow peak with an energy width below \SI{0.1}{meV} is expected to be observed at the Fermi-edge ($E_F$). In \cite{nagaoka_field_1999,nagaoka_monochromatic_1998}, a \SI{20}{meV} peak was claimed to appear near E$_F$ only at superconducting temperatures. However, we do not measure such distinct changes around T$_c$. 

Furthermore, electron-electron correlation measurements were performed resulting in no evidence of two-electron field emission from the superconducting tip, as predicted by theory \cite{yuasa_entanglement_2009}. 
This result is potentially due to two reasons. Either there is no correlated electron pair emission from the superconducting state around the Fermi energy E$_F$, or there is local heating at the beam exit area of the electrons, leading to non-superconducting conditions even at a measured overall tip temperature below T$_c$. 
A local heating mechanism that may needs to be considered is Nottingham heating \cite{vu_thien_binh_local_1992,xn_energy-exchange_nodate,bergeret_nottingham_1985,nottingham1941remarks}. The high current from a small spatial area can thereby induce localized heating that can even destabilize the structural integrity of the apex \cite{binh_field_1997}. This is due to the Nottingham effect where electrons emitted from below $E_F$ are, on average, colder than the electrons in the Fermi sea. It results in net heating, because most electrons that tunnel through the Coulomb-barrier are lower than $E_F$ at cryogenic tip temperatures \cite{nottingham1941remarks,xn_energy-exchange_nodate}. The effect was measured by Binh et al.~\cite{vu_thien_binh_local_1992} for a tungsten tip with a single atom protrusion. They observed no heating below \SI{0.3}{pA} but a local change in temperature of $\sim$\SI{30}{K} at a beam current of \SI{9}{pA} with a linear increment in between. Our count rate in Fig.~\ref{figure10} was in the fA regime and therefore significantly smaller. However, superconductors are considered to be bad thermal conductors, so our situation is potentially not comparable to normal conducting tungsten tips. This will be a matter of future investigations.

It is worth noting that the effect can be reversed if the emission mostly comes from energies above $E_F$, causing Nottingham cooling \cite{purcell_field_1994}. This could in principle be realized with a NPT by shifting the surface resonant peaks slightly above $E_F$ by lowering the extraction voltage \cite{Johnson2022,purcell_field_1994}. To actually cool efficiently at cryogenic temperatures is challenging, since there are not many electrons left above $E_F$ that can be addressed by the resonance peak. But with the possibility of precisely shifting an ultra-narrow Lorentzian excitation that is truncated at E$_F$ such as we observed in Fig.~\ref{figure5} a) and in \cite{Johnson2022}, a stable condition may be found that allows local Nottingham cooling and possibly correlated electron emission. 
This may lead to the predicted correlated and entangled two-electron emitter \cite{yuasa_entanglement_2009}. 
Alternatively, tips made from materials with a higher superconducting transition temperature could be applied, e.g.~niobium nitrite \cite{saito1999emission}. 
Such sources could become an electron-optical analog to the parametric down-conversion two-photon source in quantum optics \cite{kwiat_new_1995} and open up new modes in quantum information science, quantum metrology, and electron spectroscopy.
They could also play an important part in quantum correlation measurements in electron microscopes, such as
free-electron entanglement with optical excitations like e.g.~plasmon polaritons \cite{mechel_quantum_2021}. 

The near-monochromatic field emission of this source is highly coherent. The longitudinal coherence length is inversely proportional to the width of the energy distribution, and the transversal coherence is due to the small virtual source size. This is especially important in quantum applications with electron matter waves in e.g.~sensor technology \cite{pooch_compact_2017,rembold_vibrational_2017}, interferometry \cite{Schuetz2014}, quantum information science \cite{ropke_data_2021}, multipass-transmission electron microscopy \cite{juffmann2017multi} and quantum electron microscopy \cite{putnam_noninvasive_2009,kruit_designs_2016}. 

Our novel source could be combined with a monochromator, which would lead to an extremely low electron beam energy distribution on the meV or even sub-meV-scale, while still maintaining reasonable beam currents for microscopy applications. A high brightness free-electron source based on a Nb tip with a native energy line width on the order of tens of meV has a myriad of potential benefits throughout all electron microscopy and would open up new regimes, such as high-resolution electron energy loss spectroscopy \cite{krivanek2009high} including the direct imaging of vibrational modes \cite{krivanek2014vibrational}, the analysis of semiconductor band gaps or defects and low-loss structures in metal nanoparticles, solar cells or organic materials. The reduction in the field emitter energy distribution will also reduce chromatic aberrations, particularly in low-voltage scanning electron microscopes. This will improve their spatial resolution into the sub-nanometer range. \\

\begin{acknowledgments} 
Work at the Molecular Foundry was supported by the Office of Science, Office of Basic Energy Sciences, of the U.S. Department of Energy under Contract No.~DE-AC02-05CH11231. This material is based upon work supported by the U.S. Department of Energy, Small Business Innovation Research (SBIR) / Small Business Technology Transfer (STTR) program under Award Number DE-SC-0019675. We also acknowledge support by the Deutsche Forschungsgemeinschaft through the research grant STI 615/3-1.
\end{acknowledgments} 

\bibliography{mybib}

\end{document}